\documentclass[11pt,journal,onecolumn]{IEEEtran}
\usepackage[T1]{fontenc}
\usepackage{amsmath,amssymb,amsthm,mathtools}
\usepackage[nopatch=footnote]{microtype}
\usepackage[numbers,sort&compress]{natbib}
\usepackage[hidelinks]{hyperref}
\usepackage[nameinlink,noabbrev]{cleveref}
\usepackage{setspace}
\newtheorem{theorem}{Theorem}
\newtheorem{lemma}{Lemma}

\newtheorem{corollary}{Corollary}

\theoremstyle{definition}

\newcommand{\E}{\mathbb E}

\newcommand{\ind}{\mathbf 1}
\newcommand{\R}{\mathbb R}
\newcommand{\IF}{I_{\mathrm F}}
\newcommand{\RI}{R_{\mathrm I}}
\newcommand{\RF}{R_{\mathrm F}}
\newcommand{\bRF}{\overline R_{\mathrm F}}
\newcommand{\DCS}{D_{\mathrm{CS}}}
\DeclareMathOperator{\Var}{Var}
\DeclareMathOperator{\Cov}{Cov}
\DeclareMathOperator{\supp}{supp}

\DeclareMathOperator{\Geom}{Geom}
\DeclareMathOperator{\Exp}{Exp}
\DeclareMathOperator{\Unif}{Unif}
\allowdisplaybreaks[2]
\title{Exact Redundancy under Average and Almost-Sure Distortion Constraints}
\author{Yijun Fan}
\begin{document}
\maketitle

\begin{abstract}
We characterize the exact redundancy of lossy source
coding under average and almost-sure distortion constraints.
For general source and output alphabets, we establish
the asymptotic redundancy coefficients under suitable
regularity conditions. Under the average distortion constraint, the coefficient is half of the minimum probability that an output reveals information about the source, among all rate-distortion optimal channels. Under the almost-sure distortion constraint, the coefficient is either $1/2$ or $1$, depending on whether the $D$-tilted information is constant or not. Both converse results follow from a common change-of-measure inequality. Finally, we verify the required
conditions under both distortion requirements always hold for finite alphabets.
\end{abstract}

\section{Introduction}
Lossy source coding describes a source sequence by an output that satisfies a prescribed distortion constraint. The rate-distortion function gives the first-order communication rate, but it does not determine the excess at finite blocklength. We study this excess when distortion is bounded in expectation and when the bound holds almost surely. The latter is the $d$-semifaithful requirement.

Zhang, Yang, and Wei established logarithmic redundancy bounds for $d$-semifaithful coding~\cite{zhang1997redundancy}. For symmetric finite source-distortion pairs, Sriramu and Wagner identified the coefficient $1/2$ and showed that it remains unchanged under an average-distortion constraint~\cite{sriramu2026symmetric}. We express the coefficients for more general sources through rate-distortion optimizers and tilted information proposed in~\cite{kostina2012fixed}.
Our formulation uses functional information
\cite{GocFlamich2024,FlamichSriramuWagner2025},
which provides matching lower and upper bounds on
the expected communication cost of channel simulation up to an additive constant. It therefore serves as
a natural quantity for characterizing the communication
rate and its redundancy.

Section~\ref{sec:prelim} formulates the problem and proves the common change-of-measure inequality. Section~\ref{sec:main} states the results. Then Sections~\ref{sec:average} and~\ref{sec:hard} prove them respectively. Section~\ref{sec:finite} verifies that all regularity assumptions hold for sources with finite alphabet.

\section{Problem Formulation and Preliminaries}\label{sec:prelim}
Let $\mathcal X,\mathcal Y$ be standard Borel spaces, let $P_X$ be a probability measure on $\mathcal X$, and let $d:\mathcal X\times\mathcal Y\to[0,\infty)$ be measurable and finite-valued. The source is $X^n\sim P_X^{\otimes n}$, and
\[
d_n(x^n,y^n)=\frac1n\sum_{i=1}^n d(x_i,y_i).
\]

For $P_{XY}\ll P_X\otimes P_Y$, denote the likelihood ratio and height by
\[
r_y(x)=\frac{dP_{X|Y=y}}{dP_X}(x),\qquad
w_y(h)=P_X\{r_y(X)\ge h\},\quad h>0,
\]
respectively.
The channel simulation divergence and functional information are defined as
\begin{align}
\DCS(P_{X|Y=y}\|P_X)
&=-\int_0^\infty w_y(h)\log w_y(h)\,dh,\label{eq:cs-divergence}\\
\IF(X\to Y)&=\int \DCS(P_{X|Y=y}\|P_X)\,P_Y(dy).\label{eq:functional information}
\end{align}
We follow the convention that $0\log 0 = 0$. If the joint distribution is not absolutely continuous with respect to the product distribution, set $\IF=+\infty$. Since $0\le w_y\le1$ and $\int_0^\infty w_y(h)\,dh=1$, the divergence is nonnegative.

Define
\begin{align}
\RI(D)&=\inf_{W:\,\E d(X,Y)\le D}I(X;Y),\label{eq:ri}\\
\RF^n(D)&=\inf_{P_{Y^n|X^n}:\,\E d_n\le D}\IF(X^n\to Y^n),\label{eq:rf}\\
\bRF^n(D)&=\inf_{P_{Y^n|X^n}:\,d_n\le D\ {\rm a.s.}}\IF(X^n\to Y^n).\label{eq:hard-rf}
\end{align}
Write $\mathcal W_D$ for the set of optimizers in \eqref{eq:ri}, i.e., let
\[
\mathcal W_D =\left\{W:\mathcal X\to\mathcal P(\mathcal Y),\, \mathbb E_{P_XW}[d(X,Y)]\le D,\,
I(P_X,W)=R_I(D)
\right\}.
\] 

The coefficients of interest are the asymptotic redundancies of the average and $d$-semifaithful problems. That is, the characterization of
\begin{equation}
\alpha(D)=\lim_{n\to\infty}\frac{\RF^n(D)-n\RI(D)}{\log n},
\qquad
\bar\alpha(D)=\lim_{n\to\infty}\frac{\bRF^n(D)-n\RI(D)}{\log n}.
\label{eq:coefficients}
\end{equation}
Their existence is part of the conclusions below.

Assume that $0<R:=\RI(D)<\infty$, $\mathcal W_D$ is nonempty, and there exists an optimal output distribution $q^\star$ and $s>0$ giving
\begin{align}
j(x)&=-\log\int 2^{s(D-d(x,y))}\,q^\star(dy),
& \E j(X)&=R,\quad \E|j(X)|<\infty,\label{eq:tilted}\\
0&<\E_{P_X}2^{j(X)+s(D-d(X,y))}\le1
&&\text{for every }y\in\mathcal Y.\label{eq:dual-bound}
\end{align}
Here $j(x)$ is the information density of the optimal channel of the rate-distortion function $\RI(D)$ adjusted by a distortion term, i.e.,
\[
\imath_{X;Y^\star}(x;y)
=\log\frac{P_{Y^\star|X}(y|x)}{P_{Y^\star}(y)}
=j(x)+s(D-d(x,y)).
\]
Moreover, $j(x)$ is known as the $D$-tilted
information \cite{kostina2012fixed}. Since $s>0$, the definition of $j(x)$ implies
\[
\begin{aligned}
P_{Y^\star}\{d(x,Y^\star)\le D\}\le \E_{P_{Y^\star}}
\left[2^{s(D-d(x,Y^\star))}\right]=2^{-j(x)}.
\end{aligned}
\]
Thus, $j(x)$ provides a lower bound on the negative logarithm of the probability that a randomly generated output satisfies the distortion constraint for $x$, giving it an operational interpretation in lossy compression.

The assumptions hold automatically in the finite alphabet interior, as proved in Section~\ref{sec:finite}. Define the probability kernel
\begin{equation}
Q_y(dx)=
\frac{2^{j(x)+s(D-d(x,y))}}
{\E_{P_X}2^{j(X)+s(D-d(X,y))}}\,P_X(dx).
\label{eq:dual-q}
\end{equation}

\begin{lemma}[Rate-distortion optimality]\label{lem:dual}
Under \eqref{eq:tilted}--\eqref{eq:dual-bound}, every channel with finite mutual information and finite expected distortion satisfies
\begin{align}
I(X;Y)-R
={}&\int\left[
D(P_{X|Y=y}\|Q_y)
-\log\E_{P_X}2^{j(X)+s(D-d(X,y))}
\right]P_Y(dy)+s(D-\E d(X,Y)).
\label{eq:dual-gap}
\end{align}
A distortion-feasible channel is optimal if and only if its distortion is $D$ and, for $P_Y$-almost every $y$,
\[
P_{X|Y=y}=Q_y,\qquad
\E_{P_X}2^{j(X)+s(D-d(X,y))}=1.
\]
\end{lemma}
\begin{proof}
Taking logarithms in \eqref{eq:dual-q} and integrating against the posterior gives
\begin{align*}
D(P_{X|Y=y}\|P_X)
={}&D(P_{X|Y=y}\|Q_y)
+\E[j(X)+s(D-d(X,y))\mid Y=y]\\
&-\log\E_{P_X}2^{j(X)+s(D-d(X,y))}.
\end{align*}
The terms involving $j$ and distortion are integrable under the joint distribution. Averaging proves \eqref{eq:dual-gap}. For a feasible channel, the divergence, the negative logarithm in \eqref{eq:dual-gap}, and the distortion slack are nonnegative. Equality holds exactly when all three vanish.
\end{proof}

A key step of both the average problem and the $d$-semifaithful problem relies on the change-of-measure inequality stated below.
Here, $(t)_+=\max\{t,0\}$.
\begin{lemma}[Change of measure]\label{lem:change-measure}
Let $P\ll\nu$ be probability measures, with
\[
\frac{dP}{d\nu}=b\,2^{-F-G},
\]
where $b>0$ and $F,G$ are real-valued measurable functions. For measurable sets $A\subseteq B$ with $P(A)>0$ and $\E_P[|F|\mid A]<\infty$,
\begin{equation}
P(A)\le b\,2^{-\E_P[F\mid A]}
\E_\nu\!\left[
(F-\E_P[F\mid A]+1)_+
2^{-(F-\E_P[F\mid A])-G}\ind_B
\right].
\label{eq:change-of-measure-ineq}
\end{equation}
The expectation on the right may be infinite.
\end{lemma}
\begin{proof}
By the definition of conditional expectation,
\[
\E_P[(F-\E_P[F\mid A])\ind_A]=0.
\]
Consequently,
\begin{align*}
P(A)
&=\E_P[(F-\E_P[F\mid A]+1)\ind_A]\\
&=b\,\E_\nu[(F-\E_P[F\mid A]+1)2^{-F-G}\ind_A]\\
&\le b\,\E_\nu[(F-\E_P[F\mid A]+1)_+2^{-F-G}\ind_B]\\
&=b\,2^{-\E_P[F\mid A]}
\E_\nu[(F-\E_P[F\mid A]+1)_+
2^{-(F-\E_P[F\mid A])-G}\ind_B].
\end{align*}
The inequality uses $A\subseteq B$, positivity of the exponential, and $t\le t_+$. The signed expectation on $A$ is integrable by hypothesis; the subsequent expectation is of a nonnegative function.
\end{proof}

For a fixed reproduction block, the measure used in both converses is
\[
\nu_{y^n}=\bigotimes_{i=1}^n Q_{y_i}.
\]
Its density satisfies
\begin{equation}
\frac{dP_X^{\otimes n}}{d\nu_{y^n}}(x^n)
=
\left(\prod_{i=1}^n
\E_{P_X}2^{j(X)+s(D-d(X,y_i))}\right)
2^{-\sum_i j(x_i)-s(nD-\sum_i d(x_i,y_i))}.
\label{eq:block-density}
\end{equation}
The product in parentheses is at most one.
We will specify $F$, $G$, and $A$ in later sections for the average distortion problem and $d$-semifaithful problem.

Another common trick we use is to restrict the source to a high probability set. For any source event $A$ with $P_X(A)>0$, we have
\begin{equation}
\IF(X\to Y)\ge
P_X(A)\,\IF(X\to Y\mid A),
\label{eq:conditioning}
\end{equation}
where the conditional functional information is evaluated
under the joint distribution conditioned on $X\in A$.
To see this, let $U\sim\Unif(0,1)$ be independent of
$(X,Y)$, and put $T=Ur_Y(X)$. Then
$\IF(X\to Y)=h(T\mid Y)$. Writing
$B=\ind_{\{X\in A\}}$, the entropy scaling identity gives
\[
h(T\mid Y,B)
=\sum_b P_B(b)\IF(X\to Y\mid B=b)+I(B;Y).
\]
Since conditioning reduces differential entropy and
functional information is nonnegative,
\[
\IF(X\to Y)
\ge h(T\mid Y,B)
\ge P_X(A)\IF(X\to Y\mid A),
\]
which proves \eqref{eq:conditioning}.

Now, go back to the channel simulation problem with distortion.
An exact simulation code for $P_{Y|X}$ consists of
a random variable $Z$ independent of $X$, a measurable
encoder
\[
f:\mathcal X\times\mathcal Z\to\{0,1\}^*,
\]
and a measurable decoder
\[
g:\{0,1\}^*\times\mathcal Z\to\mathcal Y,
\]
such that $g(f(X,Z),Z)$ has conditional distribution
$P_{Y|X}$ given $X$. For $P_Z$-almost every $z$,
the codewords $\{f(x,z):x\in\mathcal X\}$ form a
prefix-free set. Let $\ell(c)$ denote the length of
a binary string $c\in\{0,1\}^*$. The expected
codeword length is $\E[\ell(f(X,Z))]$.

Let $L_n(D)$ and $\bar L_n(D)$ denote the infima
of $\E[\ell(f(X^n,Z))]$ over all exact simulation
codes satisfying $\E[d_n(X^n,Y^n)]\le D$ and
$d_n(X^n,Y^n)\le D$ almost surely, respectively,
where $Y^n=g(f(X^n,Z),Z)$. Also write
\[
\bar H_n(D)=\inf_f H(f(X^n)),
\]
where $f$ ranges over measurable, countably valued
maps satisfying $d_n(X^n,f(X^n))\le D$ almost surely.

For every exact simulation code $(Z,f,g)$,
\begin{equation}
I(X;Y)\le\IF(X\to Y)
\le \E[\ell(f(X,Z))],
\label{eq:coding-lower}
\end{equation}
where $Y=g(f(X,Z),Z)$. The first inequality and
the functional information converse hold on general
alphabets~\cite{GocFlamich2024,FlamichSriramuWagner2025}.
For a countably valued deterministic map $f$, let
$Y=f(X^n)$. For every $y$ with $P_Y(y)>0$, the
posterior distribution is
\[
P_{X^n|Y=y}
=P_X^{\otimes n}(\cdot\mid f(X^n)=y),
\]
Hence, $\frac{dP_{X^n|Y=y}}{dP_X^{\otimes n}}(x^n)
=\frac{\ind_{\{f(x^n)=y\}}}{P_Y(y)}$, implying the width function $w_y(t)=P_Y(y)\ind_{\{0<t\le 1/P_Y(y)\}}$.
This gives
\[
\DCS(P_{X^n|Y=y}\|P_X^{\otimes n}) =-\log P_Y(y).
\]
Averaging over $Y$ yields
\begin{equation}
\IF(X^n\to f(X^n))=H(f(X^n)).
\label{eq:deterministic}
\end{equation}
Consequently,
\[
\bRF^n(D)\le\bar L_n(D)\le\bar H_n(D)+1.
\]
Indeed, the first inequality follows from
\eqref{eq:coding-lower} by taking the infimum
over all channels satisfying the almost-sure
distortion constraint. For the second inequality,
any feasible deterministic map $f$ can be
implemented by encoding $f(X^n)$ using a
prefix-free Shannon code, whose expected length
is less than $H(f(X^n))+1$. Taking the infimum
over all feasible maps proves the upper bound.

If the channel density ratio is essentially bounded,
the ring-toss theorem and conditional Shannon coding
guarantee the existence of an exact simulation code
$(Z,f,g)$ satisfying
\begin{equation}
\E[\ell(f(X,Z))]
\le\IF(X\to Y)+\log e+1.
\label{eq:bounded-simulation}
\end{equation}
This additional boundedness condition belongs to the
coding theorem~\cite{HillAlajajiLinderFlamich2026}.
This restriction is not imposed on the functional information characterizations below.

\section{Characterization of the Coefficients}\label{sec:main}

\subsection{Average Distortion}
For integers $m\ge2$, let
\begin{equation}
\mathcal Y_m=
\left\{y:\ \sup_{t\in\R}
Q_y\{j(X)-s d(X,y)\in[t,t+1/m]\}
\le1-\frac1m\right\}.
\label{eq:separated-outputs}
\end{equation}
These measurable sets increase with $m$, and
\begin{equation}
\bigcup_{m\ge2}\mathcal Y_m=\{y:Q_y\ne P_X\}.
\label{eq:exhaustion}
\end{equation}
Indeed, \eqref{eq:dual-q} makes $j(X)-sd(X,y)$ constant under $Q_y$ exactly when $Q_y=P_X$. Every non-constant real distribution assigns positive probability to two separated intervals, so it belongs to the sets in \eqref{eq:separated-outputs} for sufficiently large $m$. The concentration function in \eqref{eq:separated-outputs} is a measurable function of its probability distribution.

\begin{theorem}\label{thm:average}
Assume \eqref{eq:tilted}-\eqref{eq:dual-bound} holds. Suppose that every sequence of distortion feasible channels $W_k\in \mathcal W_D$ with $X\sim P_X$, $Y_k|X\sim W_k$, and
$I(X;Y_k)\to R$, satisfies
\begin{equation}
\lim_{m\to\infty}\liminf_{k\to\infty}
P_Y^{W_k}(\mathcal Y_m)
\ge
\inf_{W\in\mathcal W_D}P_Y^W\{y:Q_y\ne P_X\}.
\label{eq:stability}
\end{equation}
Suppose also that the infimum on the right can be approached by optimizers satisfying
\begin{equation}
\E\Var\!\left(
\left.\ln\frac{dP_{X|Y}}{dP_X}(X)\right|Y
\right)<\infty.
\label{eq:average-moment}
\end{equation}
Then $\alpha(D)$ exists and
\begin{equation}
\alpha(D)=\frac12\inf_{W\in\mathcal W_D}
P_Y^W\{y:P_{X|Y=y}^W\ne P_X\}.
\label{eq:alpha-characterization}
\end{equation}
In particular,
\[
\RF^n(D)=nR+\alpha(D)\log n+o(\log n).
\]
\end{theorem}

Condition \eqref{eq:stability} controls nearly optimal channels at each fixed concentration scale. It allows informative tilted distributions with arbitrarily small spread. Condition \eqref{eq:average-moment} is an average conditional-variance condition on the optimizers used for achievability.

There is a familiar sufficient way to verify \eqref{eq:stability}. Suppose $\mathcal Y$ is Polish, the distribution of $j(X)-sd(X,y)$ under $Q_y$ depends weakly continuously on $y$, and every nearly optimal feasible sequence has a subsequence of output distributions converging weakly to the output distribution of an optimizer. Then \eqref{eq:stability} follows. A proof is given after Theorem~\ref{thm:average}. 

The formula gives $0\le\alpha(D)\le1/2$. If the infimum is attained, then $\alpha(D)>0$, since an optimizer with posterior $P_X$ almost surely would have mutual information zero, contradicting to the optimality the infimum with $I(X;Y)=R>0$. 

\subsection{Almost Sure Distortion}
Write
\[
V(D)=\Var(j(X))\in[0,\infty].
\]
Here $V(D)$ may be infinite. 
Assume that there exists a sequence of measurable
sets $G_n\subseteq\mathcal X^n$ satisfying the
following conditions for all sufficiently large $n$.
The assumptions below specify the concentration and covering estimates needed on such set $G_n$.

Assume that on $G_n^c$, there is a measurable prefix code whose reconstruction satisfies $d_n\le D$ almost surely. If its length is $\ell_n(X^n)$, require
\begin{equation}
\E\!\left[
\left(\left|\sum_{i=1}^n j(X_i)\right|+\log n+\ell_n(X^n)\right)
\ind_{\{X^n\notin G_n\}}
\right]=O(1).
\label{eq:exceptional-cost}
\end{equation}
Only the contribution from $G_n^c$ is constrained and the code need not be efficient elsewhere. This condition implies $P_X^{\otimes n}(G_n)\to1$.

For the fixed optimal output distribution $q^\star$ in \eqref{eq:tilted}, put
\begin{equation}
p_n(x^n)=(q^\star)^{\otimes n}\{y^n:d_n(x^n,y^n)\le D\}.
\label{eq:ball-prob}
\end{equation}
Require $p_n>0$ almost surely on $G_n$ and
\begin{align}
\E[-\log p_n(X^n)\,\ind_{\{X^n\in G_n\}}]
&\le nR+\tfrac12\log n+O(1),\label{eq:covering-mean}\\
\Var(\log p_n(X^n)\mid X^n\in G_n)
&=
\begin{cases}
O(1),&V(D)=0,\\
O(n),&V(D)>0.
\end{cases}\label{eq:covering-variance}
\end{align}
Finally, for each $y^n\in\mathcal Y^n$, recall that $\nu_{y^n}=\bigotimes_{i=1}^n Q_{y_i}$,
where $Q_y$ is defined in \eqref{eq:dual-q}.
Assume there exists a constant $C$, independent of
$n$, $y^n$, $u$, and $v$, such that for all sufficiently
large $n$, all $y^n\in\mathcal Y^n$, $u\in\R$, and $v\ge0$,
\begin{equation}
\nu_{y^n}\!\left\{
\begin{array}{l}
X^n\in G_n,\quad \sum_i j(X_i)\in[u,u+1),\\
nD-\sum_i d(X_i,y_i)\in[v,v+1)
\end{array}\right\}
\le
\begin{cases}
C/\sqrt n,&V(D)=0,\\
C/n,&V(D)>0.
\end{cases}
\label{eq:hard-local}
\end{equation}
When $V(D)=0$, the first sum is constant, so this is a scalar concentration condition. When $V(D)>0$, it is a two-dimensional local concentration condition.

\begin{theorem}\label{thm:hard}
Assume \eqref{eq:tilted}--\eqref{eq:dual-bound} and \eqref{eq:exceptional-cost}--\eqref{eq:hard-local} holds. Then $\bar\alpha(D)$ exists and
\begin{equation}
\bar\alpha(D)=\frac{1+\ind_{\{V(D)>0\}}}{2}.
\label{eq:hard-characterization}
\end{equation}
Moreover, each of $\bRF^n(D)$, $\bar H_n(D)$, and $\bar L_n(D)$ equals
\begin{equation}
nR+\frac{1+\ind_{\{V(D)>0\}}}{2}\log n+O(1).
\label{eq:hard-expansion}
\end{equation}
\end{theorem}

The conditions in Theorem~\ref{thm:hard} depend only on
the source distribution and the fixed measures $q^\star $ and $\{Q_y\}_{y\in\mathcal Y}$ determined
by the single-letter rate-distortion problem. In particular, they do not require knowledge of an optimal block channel. 
Condition \eqref{eq:exceptional-cost} ensures that source
sequences outside $G_n$ can be reproduced with negligible expected coding cost, and their probability
being small is not sufficient unless the cost of handling them is also controlled. Conditions
\eqref{eq:covering-mean} and
\eqref{eq:covering-variance} control the mean and
fluctuations of the logarithmic distortion-ball
probabilities needed for achievability, while
\eqref{eq:hard-local} provides the uniform local
concentration bound needed for the converse.
For finite alphabets, all these conditions are satisfied. 
For general alphabets, they must be verified separately,
since a central limit theorem alone does not
guarantee the fixed-width concentration bound
required in \eqref{eq:hard-local}.

\section{Proof of the Average-Distortion Characterization}
\label{sec:average}
\subsection{Product Channels and Concentration Bound}
\begin{lemma}[Product-channel upper bound]\label{lem:product}
For any channel on standard Borel spaces with $I(X;Y)<\infty$, put
\[
J=\ln\frac{dP_{X|Y}}{dP_X}(X),\qquad
\sigma_c^2=\E\Var(J\mid Y).
\]
If $\sigma_c^2<\infty$, then
\begin{equation}
0\le\IF(X^n\to Y^n)-nI(X;Y)
\le\frac12\log\bigl(2\pi e(1+n\sigma_c^2)\bigr)-\log e.
\label{eq:product-bound}
\end{equation}
\end{lemma}
\begin{proof}
Let $U\sim\Unif(0,1)$ be independent of $(X,Y)$. Since $Ur_Y(X)$ has conditional density $w_Y$, we have
\[
\IF(X\to Y)=h(Ur_Y(X)\mid Y).
\]
With $E=-\ln U\sim\Exp(1)$, the entropy change of variables gives
\begin{equation}
\IF(X\to Y)-I(X;Y)=h(J-E\mid Y)-\log e.
\label{eq:entropy-identity}
\end{equation}
The conditional entropy on the right is well defined. It is at least $h(-E)=\log e$, and its average upper bound is finite by the conditional-variance hypothesis and the Gaussian entropy bound.

For the product channel there is one exponential variable for the whole block. Conditional on $Y^n$, the $J_i$ are independent, and
\[
\Var\left(\sum_iJ_i-E\ \middle|\ Y^n\right)
=1+\sum_i\Var(J_i\mid Y_i).
\]
Consequently,
\begin{align*}
\IF(X^n\to Y^n)-nI(X;Y)
&=h\left(\sum_iJ_i-E\ \middle|\ Y^n\right)-\log e\\
&\le\frac12\E\log\left(2\pi e\left[1+\sum_i\Var(J_i\mid Y_i)\right]\right)-\log e\\
&\le\frac12\log\bigl(2\pi e(1+n\sigma_c^2)\bigr)-\log e.
\end{align*}
The lower bound follows from \eqref{eq:coding-lower}.
\end{proof}

\begin{lemma}\label{lem:scalar-converse}
For every fixed $m\ge2$, there is $C_m<\infty$ such that every average-distortion feasible block channel satisfies
\begin{equation}
\IF(X^n\to Y^n)\ge nR+
\frac12\E\log\left(1+\sum_i\ind_{\{Y_i\in\mathcal Y_m\}}\right)-C_m.
\label{eq:uniform-average-converse}
\end{equation}
\end{lemma}
\begin{proof}
For fixed $y^n$, write
\[
S(x^n,y^n)
=\sum_{i=1}^n[j(x_i)+s(D-d(x_i,y_i))],
\qquad
K=\sum_{i=1}^n\ind_{\{y_i\in\mathcal Y_m\}}.
\]
For notation simplicity, write $S(x^n)=S(x^n,y^n)$
when $y^n$ is fixed.
Under $\nu_{y^n}$, the summands are independent. For each coordinate $i$ with $y_i\in\mathcal Y_m$, \eqref{eq:separated-outputs} implies
\[
\sup_{t\in\R}
Q_{y_i}\{j(X)+s(D-d(X,y_i))
\in[t,t+1/m]\}
\le 1-\frac1m,
\]
since the addition of the constant $sD$ does not
affect concentration. Thus, for each of these $K$ coordinates, every interval of length $1/m$ excludes at least $1/m$ of the probability mass.
Applying the Kolmogorov--Rogozin inequality
~\cite{rogozin1961increase} to these independent
summands gives, for $K\ge1$,
\[
\sup_t\nu_{y^n}\{S\in[t,t+1)\}
\le
\frac{C}{\sqrt{K(1/m)^2(1/m)}}
=\frac{Cm^{3/2}}{\sqrt K}.
\]
Since $m$ is fixed,
and the case $K=0$ is trivial, we obtain
\[
\sup_t\nu_{y^n}\{S\in[t,t+1)\}
\le\frac{C_m}{\sqrt{K+1}}.
\] 

For a set $A$ of positive source probability on which $S$ is integrable, put $T=S-\E[S\mid A]$. Apply Lemma~\ref{lem:change-measure} with $F=S$, $G=0$, and $B=\mathcal X^n$. The prefactor in \eqref{eq:block-density} is at most one, so
\[
P_X^{\otimes n}(A)\le
2^{-\E[S\mid A]}\E_{\nu_{y^n}}[(T+1)_+2^{-T}].
\]
The integrand vanishes when $T\le-1$. For each integer $k\ge0$, if $T\in[k-1,k)$, then $(T+1)_+\le k+1$ and $2^{-T}\le 2^{1-k}$. Hence, $(T+1)_+2^{-T}\le(k+1)2^{1-k}$.
Moreover, since $T$ differs from $S$ only by a constant, the preceding concentration bound gives
\[
\nu_{y^n}\{T\in[k-1,k)\}
\le\frac{C_m}{\sqrt{K+1}}.
\]
Consequently,
\begin{align*}
\E_{\nu_{y^n}}[(T+1)_+2^{-T}]&\le
\sum_{k=0}^\infty(k+1)2^{1-k}
\nu_{y^n}\{T\in[k-1,k)\}\\
&\le
\frac{C_m}{\sqrt{K+1}}
\sum_{k=0}^\infty(k+1)2^{1-k}\\
&=
\frac{8C_m}{\sqrt{K+1}}.
\end{align*}
Absorbing the factor $8$ into $C_m$, we obtain
\begin{equation}
P_X^{\otimes n}(A)\le
\frac{C_m}{\sqrt{K+1}}\,2^{-\E[S\mid A]}.
\label{eq:scalar-set}
\end{equation}

It then suffices to consider channels with finite functional information. At a fixed output, let $r=dP_{X^n|y^n}/dP_X^{\otimes n}$, $A_h=\{r\ge h\}$, and $w(h)=P_X^{\otimes n}(A_h)$. For almost every output, $S$ is integrable under its posterior, because
\[
\E|S(X^n,Y^n)|\le n\E|j(X)|+s(nD+\E\textstyle\sum_i d(X_i,Y_i))<\infty.
\]
For $h>0$, integrability on $A_h$ follows from
$\int_{A_h}|S|\,dP_X^{\otimes n}\le h^{-1}\int |S|\,dP_{X^n|y^n}$.
Tonelli's theorem, followed by Fubini's theorem for the integrable signed term, gives
\[
\int_0^\infty w(h)\,dh=1,\qquad
\int_0^\infty w(h)\E[S\mid A_h]\,dh=\E[S\mid Y^n=y^n].
\]
Applying \eqref{eq:scalar-set} to $A=A_h$ and
taking negative logarithms, when $w(h)>0$, we obtain
\[
-\log w(h)
\ge \E[S\mid A_h]
+\frac12\log(K+1)-\log C_m.
\]
Multiplying by $w(h)$ and integrating over $h>0$
gives
\begin{align*}
\DCS(P_{X^n|y^n}\|P_X^{\otimes n})&=-\int_0^\infty w(h)\log w(h)\,dh\\
&\ge
\int_0^\infty w(h)\E[S\mid A_h]\,dh +\frac12\log(K+1)-\log C_m\\
&=\E[S\mid Y^n=y^n] +\frac12\log(K+1)-\log C_m.
\end{align*}
Averaging over $Y^n$ and using the definition of functional information yields
\begin{align*}
\IF(X^n\to Y^n)
&\ge \E[S(X^n,Y^n)]
+\frac12\E\log(K+1)-\log C_m\\
&\ge nR+\frac12\E\log(K+1)-\log C_m,
\end{align*}
where the last inequality follows from
\begin{align*}
\E[S(X^n,Y^n)]=nR+s\left(
nD-\E\sum_{i=1}^n d(X_i,Y_i)
\right) \ge nR
\end{align*}
by the average-distortion constraint.
Recalling the definition of $K$ and absorbing
$\log C_m$ into the constant proves
\eqref{eq:uniform-average-converse}.
\end{proof}

\subsection{Proof of Theorem~\ref{thm:average}}
\begin{proof}
We first prove the upper bound.
Fix an optimizer $W$ satisfying \eqref{eq:average-moment}, and write $\theta=P_Y\{y:Q_y\ne P_X\}$.
Here $\theta$ is the informative output probability of this fixed channel. Let $B=\ind_{\{Q_Y\ne P_X\}}$. 
By Lemma~\ref{lem:dual}, every uninformative posterior equals $P_X$. Hence, for every measurable set $A\subseteq\mathcal X$,
\[
\begin{aligned}
\Pr\{X\in A,B=0\} =(1-\theta)P_X(A),\qquad \Pr\{X\in A,B=1\} =\theta P_X(A).
\end{aligned}
\]
Moreover, $X$ and $Y$ are conditionally independent given
$B=0$, since $P_{X|Y}=P_X$ on this event.
Since $B$ is a function of $Y$, the chain rule for mutual information gives
\[
\begin{aligned}
R=I(X;Y) =I(X;B)+I(X;Y\mid B) =\theta I(X;Y\mid B=1).
\end{aligned}
\]
In particular, $\theta>0$ by $R>0$.

Draw one $B$ for the entire source block. Conditional on it, use the corresponding conditional channel independently at all coordinates. Its expected distortion is still $D$. The two branches of $B$ have disjoint output supports, the uninformative branch with $B=0$ has functional information zero, and
\[
\IF(X^n\to Y^n)=\theta\,\IF(X^n\to Y^n\mid B=1).
\] 
Let \[
\sigma_1^2
=\E\left[
\left.
\Var\left(
\left.
\ln\frac{dP_{X|Y}}{dP_X}(X)
\right|Y
\right)
\right|B=1
\right]<\infty.
\]
Since the uninformative branch has zero functional
information, Lemma~\ref{lem:product} gives
\begin{align*}
\RF^n(D)
&\le \IF(X^n\to Y^n)\\
&=\theta\IF(X^n\to Y^n\mid B=1)\\
&\le \theta\left[
nI(X;Y\mid B=1)
+\frac12\log\bigl(2\pi e(1+n\sigma_1^2)\bigr)
-\log e
\right]\\
&=nR+\frac{\theta}{2}
\log\bigl(2\pi e(1+n\sigma_1^2)\bigr)
-\theta\log e\\
&\le nR+\frac{\theta}{2}\log n+C,
\end{align*}
where
\[
C=\frac{\theta}{2}
\log\bigl(2\pi e(1+\sigma_1^2)\bigr)
-\theta\log e
\]
is independent of $n$. The last inequality follows
from $1+n\sigma_1^2\le n(1+\sigma_1^2)$ for $n\ge1$.
Taking the upper limit after division by $\log n$, and then taking the infimum over the permitted optimizers, proves the required upper bound on the coefficient. 

For the converse, select a feasible block channel satisfying
\[
\IF(X^n\to Y^n)\le\RF^n(D)+1.
\]
Let $W_{n,i}$ be its coordinate channels and set $\widetilde W_n=n^{-1}\sum_{i=1}^n W_{n,i}$.
The averaged channel $\widetilde W_n$ satisfies
the single-letter average-distortion constraint.
Let $(\widetilde X_n,\widetilde Y_n)\sim P_X\widetilde W_n$. 
By convexity of mutual information and the independence of the source coordinates,
\[
R\le I(\widetilde X_n;\widetilde Y_n)
\le\frac1n\sum_{i=1}^n I(X_i;Y_i)
\le\frac1n I(X^n;Y^n)
\le\frac1n\IF(X^n\to Y^n).
\]
Moreover, the preceding achievability bound gives
\[
\IF(X^n\to Y^n)
\le\RF^n(D)+1
\le nR+\frac{\theta}{2}\log n+C+1,
\]
where $C$ is independent of $n$. Therefore,
\begin{equation}
I(\widetilde X_n;\widetilde Y_n)\to R.
\label{eq:coordinate-stability}
\end{equation}
Thus, the sequence $\widetilde W_n$ satisfies
the assumptions of \eqref{eq:stability}.

For each fixed $m\ge2$, let
\[
K=\sum_{i=1}^n\ind_{\{Y_i\in\mathcal Y_m\}}.
\]
Since $0\le K\le n$, the concavity of the logarithm gives $\log(K+1)\ge\frac{K}{n}\log(n+1)$.
Moreover, by the definition of the averaged channel
$\widetilde W_n$, we have
\begin{align*}
\frac{\E K}{n}=\frac1n\sum_{i=1}^n
\Pr\{Y_i\in\mathcal Y_m\}=P_Y^{\widetilde W_n}(\mathcal Y_m).
\end{align*}
Applying Lemma~\ref{lem:scalar-converse} to the
selected block channel, we obtain
\begin{align*}
\RF^n(D)+1
&\ge\IF(X^n\to Y^n)\\
&\ge nR+\frac12\E\log(K+1)-C_m\\
&\ge nR+\frac{\log(n+1)}{2n}\E K-C_m\\
&=nR+\frac12\log(n+1)
P_Y^{\widetilde W_n}(\mathcal Y_m)-C_m.
\end{align*}
Consequently,
\[
\frac{\RF^n(D)-nR}{\log n}
\ge
\frac{\log(n+1)}{2\log n}
P_Y^{\widetilde W_n}(\mathcal Y_m)
-\frac{C_m+1}{\log n}.
\]
Since $m$ is fixed, $C_m$ is independent of $n$.
Using $\log(n+1)/\log n\to1$ and
$0\le P_Y^{\widetilde W_n}(\mathcal Y_m)\le1$,
we conclude that
\begin{equation}
\liminf_{n\to\infty}
\frac{\RF^n(D)-nR}{\log n}
\ge
\frac12\liminf_{n\to\infty}
P_Y^{\widetilde W_n}(\mathcal Y_m).
\end{equation}
Since this holds for every fixed $m\ge2$, letting
$m\to\infty$ and applying \eqref{eq:stability} yields
\[
\begin{aligned}
\liminf_{n\to\infty}
\frac{\RF^n(D)-nR}{\log n}
&\ge
\frac12\lim_{m\to\infty}
\liminf_{n\to\infty}
P_Y^{\widetilde W_n}(\mathcal Y_m)\\
&\ge
\frac12\inf_{W\in\mathcal W_D}
P_Y^W\{y:Q_y\ne P_X\}\\
&=
\frac12\inf_{W\in\mathcal W_D}
P_Y^W\{y:P_{X|Y=y}^W\ne P_X\},
\end{aligned}
\]
where the last equality follows from
Lemma~\ref{lem:dual} since $Q_y=P_{X|Y=y}$ almost surely at an optimizer. This proves the lower bound in \eqref{eq:alpha-characterization}, and hence the existence and value of the limit.
\end{proof}


\section{Proof of the Almost-Sure Characterization}\label{sec:hard}
\subsection{Achievability by Random Covering}
Draw an independent countable codebook of reproduction blocks with distribution $(q^\star)^{\otimes n}$. 
Let $G_n\subseteq\mathcal X^n$ be the measurable
source sets satisfying
\eqref{eq:exceptional-cost}--\eqref{eq:hard-local}.
Draw an independent countable codebook
$\{Y^n(j)\}_{j\ge1}$, whose codewords are i.i.d.
according to $(q^\star)^{\otimes n}$.
For $X^n\in G_n$, define
\[
J=\min\{j\ge1:d_n(X^n,Y^n(j))\le D\}.
\]
Recall that $p_n(x^n)=(q^\star)^{\otimes n}\{y^n:d_n(x^n,y^n)\le D\}$ by \eqref{eq:ball-prob}.
Since $p_n(x^n)>0$ for
$P_X^{\otimes n}$-almost every $x^n\in G_n$,
the index $J$ is finite almost surely and
\[
J\mid X^n=x^n\sim\Geom(p_n(x^n)).
\]
For $X^n\notin G_n$, use the auxiliary prefix code assumed in \eqref{eq:exceptional-cost}, which produces a reconstruction satisfying $d_n(X^n,Y^n)\le D$ almost surely and has description length $\ell_n(X^n)$.
Jensen's inequality gives $\E[\log J\mid X^n]\le-\log p_n(X^n)$. The geometric distribution also obeys, uniformly over $0<p\le1$,
\[
\Pr\{pJ\le e^{-v}\}\le e^{-v},\qquad
\Pr\{pJ\ge e^v\}\le e^{1-e^v},\quad v\ge0.
\]
Integrating these tail bounds gives $\sup_{0<p\le1}\E[(\log J+\log p)^2]<\infty$.

Since $J\mid X^n=x^n$ is geometrically distributed
with parameter $p_n(x^n)$, the preceding uniform
second-moment bound gives
\[
\E\left[
(\log J+\log p_n(X^n))^2
\mid X^n\in G_n
\right]\le C
\]
for some constant $C<\infty$ independent of $n$.
Consequently,
\begin{align*}
\Var(\log J\mid X^n\in G_n)&\le
2\Var(\log p_n(X^n)\mid X^n\in G_n)+
2\E\left[
(\log J+\log p_n(X^n))^2
\mid X^n\in G_n
\right]\\
&\le
2\Var(\log p_n(X^n)\mid X^n\in G_n)+2C.
\end{align*}
Combining this with \eqref{eq:covering-variance},
we obtain
\begin{equation}
\Var(\log J\mid X^n\in G_n)=
\begin{cases}
O(1),&V(D)=0,\\
O(n),&V(D)>0.
\end{cases}
\label{eq:index-variance}
\end{equation}

For a positive integer $J$ with finite $\E\log J$ and $\Var(\log J)$, take independent $U\sim\Unif(0,1)$. Since $h(J+U)=H(J)$, a logarithmic change of variables and the Gaussian entropy bound give
\begin{align}
H(J)=h(\ln(J+U))+\E\log(J+U) \le \E\log J+\tfrac12\log(1+\Var(\log J))+C,
\label{eq:integer-entropy}
\end{align}
To prove the last line, observe that $0\le\ln(J+U)-\ln J\le\ln2$, and hence
\[
\Var(\ln(J+U))\le2\Var(\ln J)+2(\ln2)^2,\qquad
\E\log(J+U)\le\E\log J+1.
\]

On $G_n^c$, use the reconstruction given in \eqref{eq:exceptional-cost}. Let $\mathcal C$ denote the random codebook and let $Y^n$ be the resulting reconstruction. Let $B=\ind_{\{X^n\in G_n\}}$. Combining
\eqref{eq:integer-entropy}, \eqref{eq:index-variance},
\eqref{eq:exceptional-cost}, and
\eqref{eq:covering-mean}, we obtain
\begin{align*}
&H(Y^n\mid\mathcal C)\\
&\le H(B,Y^n\mid\mathcal C)\\
&=H(B) +\Pr(B=1)H(Y^n\mid B=1,\mathcal C)+\Pr(B=0)H(Y^n\mid B=0,\mathcal C)\\
&\le 1+\Pr(B=1)H(J\mid B=1,\mathcal C)+\E[\ell_n(X^n)\ind_{\{B=0\}}]\\
&\le 1+\Pr(B=1)H(J\mid B=1)+\E[\ell_n(X^n)\ind_{\{B=0\}}]\\
&\le 1+\Pr(B=1)\E[\log J\mid B=1]+\frac{\Pr(B=1)}{2}
\log\bigl(1+\Var(\log J\mid B=1)\bigr)+C_0+\E[\ell_n(X^n)\ind_{\{B=0\}}]\\
&\le 1+\E[-\log p_n(X^n)\ind_{\{B=1\}}]+\frac12
\log\bigl(1+C_1n^{\ind_{\{V(D)>0\}}}\bigr)+C_0+\E[\ell_n(X^n)\ind_{\{B=0\}}]\\
&\le \E[-\log p_n(X^n)\ind_{\{B=1\}}]+\frac{\ind_{\{V(D)>0\}}}{2}\log n+O(1)\\
&\le nR+\frac12\log n
+\frac{\ind_{\{V(D)>0\}}}{2}\log n+O(1)\\
&=nR+\frac{1+\ind_{\{V(D)>0\}}}{2}
\log n+O(1),
\end{align*}
where $C_0$ and $C_1$ are finite constants
independent of $n$.

For every $x^n\in G_n$ with $p_n(x^n)>0$,
the probability that none of the first $k$
codewords satisfies the distortion constraint is $\Pr\{J>k\mid X^n=x^n\}=(1-p_n(x^n))^k.$
Consequently,
\[
\Pr\{J=\infty\mid X^n=x^n\}
=\lim_{k\to\infty}(1-p_n(x^n))^k=0.
\]
Since $p_n(x^n)>0$ for
$P_X^{\otimes n}$-almost every $x^n\in G_n$,
Tonelli's theorem gives
\begin{align*}
0=\int_{G_n}
   \Pr\{J=\infty\mid X^n=x^n\}
   P_X^{\otimes n}(dx^n)=\int
   P_X^{\otimes n}
   \{x^n\in G_n:J_{C}(x^n)=\infty\}
   P_{\mathcal C}(dC).
\end{align*}
Thus, for $P_{\mathcal C}$-almost every
codebook $C$, the first successful
index $J_{C}(x^n)$ is finite for
$P_X^{\otimes n}$-almost every $x^n\in G_n$.

For each such fixed codebook, the random covering construction on $G_n$, together with the auxiliary prefix code on $G_n^c$, defines a measurable, countably valued quantizer whose reconstruction satisfies
$d_n(X^n,Y^n)\le D$ almost surely.
By the definition of $\bar H_n(D)$,
\begin{align*}
\bar H_n(D) &\le \inf_{ C:\,
    P_X^{\otimes n}(J_{ C}=\infty,\,
    X^n\in G_n)=0}
    H(Y^n\mid\mathcal C= C)\\
&\le \int H(Y^n\mid\mathcal C= C)
    P_{\mathcal C}(d C)\\
&=H(Y^n\mid\mathcal C)\\
&\le nR+
\frac{1+\ind_{\{V(D)>0\}}}{2}\log n+O(1).
\end{align*}

\subsection{Converse from the Change-of-Measure Inequality}

Let $G_n\subseteq\mathcal X^n$ be the measurable
source sets satisfying
\eqref{eq:exceptional-cost}--\eqref{eq:hard-local}.
In particular, \eqref{eq:exceptional-cost} implies
$P_X^{\otimes n}(G_n)\to1$, while
\eqref{eq:hard-local} provides the required
concentration bound on $G_n$.

Fix $y^n\in\mathcal Y^n$ and a measurable set $A\subseteq
G_n\cap\{x^n:d_n(x^n,y^n)\le D\}$
with $P_X^{\otimes n}(A)>0$.
Applying Lemma~\ref{lem:change-measure} and
\eqref{eq:block-density} with
\[
F=\sum_{i=1}^n j(X_i),\quad
G=s\left(nD-\sum_{i=1}^n d(X_i,y_i)\right), \quad B=G_n\cap\{x^n:d_n(x^n,y^n)\le D\},
\]
we obtain
\begin{align*}
P_X^{\otimes n}(A)
&\le
2^{-\E[\sum_i j(X_i)\mid X^n\in A]}\\
&\quad\times
\E_{\nu_{y^n}}\Bigg[
\left(
\sum_i j(X_i)
-\E\left[\sum_i j(X_i)\mid X^n\in A\right]
+1
\right)_+\times
2^{-\sum_i j(X_i)
+\E[\sum_i j(X_i)\mid X^n\in A]}\\
&\qquad\qquad\times
2^{-s(nD-\sum_i d(X_i,y_i))}
\ind_{\{X^n\in G_n,\,
d_n(X^n,y^n)\le D\}}
\Bigg].
\end{align*}

Partition the support of the integrand into unit
rectangles according to
\[
\sum_i j(X_i)
-\E\left[\sum_i j(X_i)\mid X^n\in A\right]
\in[k-1,k)
\]
and
\[
nD-\sum_i d(X_i,y_i)\in[l,l+1),
\qquad k,l\ge0.
\]
On each rectangle, the integrand is bounded by
$(k+1)2^{1-k-sl}$, while
\eqref{eq:hard-local} bounds the probability of
the rectangle under $\nu_{y^n}$ by
$Cn^{-(1+\ind_{\{V(D)>0\}})/2}$, uniformly
over $k$, $l$, $y^n$, and $A$.
Summing these bounds over all rectangles gives
\begin{align*}
P_X^{\otimes n}(A)
&\le
\frac{C\,2^{-\E[\sum_i j(X_i)\mid X^n\in A]}}
{n^{(1+\ind_{\{V(D)>0\}})/2}}
\sum_{k,l\ge0}(k+1)2^{1-k-sl}\\
&=
\frac{8C}{1-2^{-s}}\,
n^{-(1+\ind_{\{V(D)>0\}})/2}\times
2^{-\E[\sum_i j(X_i)\mid X^n\in A]}.
\end{align*}
Since $A\subseteq G_n$, it follows that
\begin{align*}
-\log P_X^{\otimes n}(A\mid G_n)
&=-\log P_X^{\otimes n}(A)
+\log P_X^{\otimes n}(G_n)\\
&\ge
\E\left[\sum_i j(X_i)\mid X^n\in A\right]+
\frac{1+\ind_{\{V(D)>0\}}}{2}\log n+
\log P_X^{\otimes n}(G_n)
-\log\frac{8C}{1-2^{-s}}.
\end{align*}

Now consider any block channel satisfying
$d_n(X^n,Y^n)\le D$ almost surely and having
finite functional information.
Condition its joint distribution on
$X^n\in G_n$.
For $P_{Y^n|X^n\in G_n}$-almost every $y^n$,
the conditional posterior is supported on
\[
G_n\cap\{x^n:d_n(x^n,y^n)\le D\}.
\]
For each $h>0$, apply the preceding set bound to
the likelihood-ratio superlevel set
\[
A=
\left\{
x^n\in G_n:
\frac{
dP_{X^n|Y^n=y^n,X^n\in G_n}
}{
dP_X^{\otimes n}(\cdot\mid G_n)
}(x^n)\ge h
\right\}.
\]

By the definition of channel simulation divergence,
we obtain
\begin{align*}
&\DCS\left(
P_{X^n|Y^n=y^n,X^n\in G_n}
\middle\|
P_X^{\otimes n}(\cdot\mid G_n)
\right)\\
&\quad=
-\int_0^\infty
P_X^{\otimes n}(A\mid G_n)
\log P_X^{\otimes n}(A\mid G_n)\,dh\\
&\quad\ge
\int_0^\infty
P_X^{\otimes n}(A\mid G_n)
\E\left[
\sum_i j(X_i)\mid X^n\in A
\right]dh\\
&\qquad+
\left(
\frac{1+\ind_{\{V(D)>0\}}}{2}\log n
+\log P_X^{\otimes n}(G_n)
-\log\frac{8C}{1-2^{-s}}
\right)\times
\int_0^\infty
P_X^{\otimes n}(A\mid G_n)\,dh\\
&\quad=
\E\left[
\sum_i j(X_i)
\mid Y^n=y^n,\ X^n\in G_n
\right]+
\frac{1+\ind_{\{V(D)>0\}}}{2}\log n
+\log P_X^{\otimes n}(G_n)
-\log\frac{8C}{1-2^{-s}},
\end{align*}
where the last equality holds by $\int_0^\infty
P_X^{\otimes n}(A\mid G_n)\,dh=1$ and, by Fubini's theorem,
\begin{align*}
&\int_0^\infty
P_X^{\otimes n}(A\mid G_n)
\E\left[
\sum_i j(X_i)\mid X^n\in A
\right]dh=
\E\left[
\sum_i j(X_i)
\mid Y^n=y^n,\ X^n\in G_n
\right].
\end{align*}

Averaging the divergence bound over
$P_{Y^n|X^n\in G_n}$ gives
\begin{align*}
\IF(X^n\to Y^n\mid X^n\in G_n)&\quad=
\int
\DCS\left(
P_{X^n|Y^n=y^n,X^n\in G_n}
\middle\|
P_X^{\otimes n}(\cdot\mid G_n)
\right)
P_{Y^n|X^n\in G_n}(dy^n)\\
&\quad\ge
\E\left[
\sum_i j(X_i)\mid X^n\in G_n
\right]\\
&\qquad+
\frac{1+\ind_{\{V(D)>0\}}}{2}\log n
+\log P_X^{\otimes n}(G_n)
-\log\frac{8C}{1-2^{-s}}.
\end{align*}

Finally, applying \eqref{eq:conditioning} and using $\E j(X)=R$, we obtain
\begin{align*}
&\IF(X^n\to Y^n)\\
&\ge
P_X^{\otimes n}(G_n)
\IF(X^n\to Y^n\mid X^n\in G_n)\\
&\ge
\E\left[
\sum_i j(X_i)
\ind_{\{X^n\in G_n\}}
\right]+
\frac{1+\ind_{\{V(D)>0\}}}{2}
P_X^{\otimes n}(G_n)\log n+
P_X^{\otimes n}(G_n)
\log P_X^{\otimes n}(G_n)-
P_X^{\otimes n}(G_n)
\log\frac{8C}{1-2^{-s}}\\
&=
nR-
\E\left[
\sum_i j(X_i)
\ind_{\{X^n\notin G_n\}}
\right]+
\frac{1+\ind_{\{V(D)>0\}}}{2}\log n-
\frac{1+\ind_{\{V(D)>0\}}}{2}
P_X^{\otimes n}(G_n^c)\log n\\
&\quad+
P_X^{\otimes n}(G_n)
\log P_X^{\otimes n}(G_n)-
P_X^{\otimes n}(G_n)
\log\frac{8C}{1-2^{-s}}\\
&\ge
nR+\frac{1+\ind_{\{V(D)>0\}}}{2}
\log n-O(1).
\end{align*}
The last inequality follows from
\eqref{eq:exceptional-cost}, which ensures
\[
\E\left[
\left|\sum_i j(X_i)\right|
\ind_{\{X^n\notin G_n\}}
\right]=O(1)
\]
and
\[
P_X^{\otimes n}(G_n^c)\log n=O(1).
\]
Moreover,
$P_X^{\otimes n}(G_n)\log P_X^{\otimes n}(G_n)$
is uniformly bounded, and all other constants
are independent of $n$.

Since the converse holds for every almost-surely
distortion-feasible block channel, taking the
infimum yields
\[
\bRF^n(D)
\ge
nR+\frac{1+\ind_{\{V(D)>0\}}}{2}
\log n-O(1).
\]
Together with the achievability bound, this proves
\eqref{eq:hard-expansion} and hence Theorem~\ref{thm:hard}.

\section{Finite-Alphabet Specialization}\label{sec:finite}
We now verify the assumptions of Theorems~\ref{thm:average} and \ref{thm:hard} for finite source. 
The first subsection establishes the tilted representation \eqref{eq:tilted}--\eqref{eq:dual-bound}, the stability condition \eqref{eq:stability}, and the moment condition \eqref{eq:average-moment}. 
The second verifies the exceptional-block and covering conditions \eqref{eq:exceptional-cost}--
\eqref{eq:covering-variance}.
The third proves the local concentration
condition \eqref{eq:hard-local}. 

\begin{corollary}\label{cor:finite}
Suppose $\mathcal X,\mathcal Y$ are finite, $P_X$ has full support, and $d$ is finite-valued. Write
\[
D_{\min}=\E\min_y d(X,y),\qquad D_0=\min_y\E d(X,y).
\]
For $D_{\min}<D<D_0$,
\[
\alpha(D)=\frac12\min_{W\in\mathcal W_D}
P_Y^W\{y:P_{X|Y=y}^W\ne P_X\}.
\]
If additionally
\begin{equation}
D\ge\max_x\min_y d(x,y),
\label{eq:hard-feasible}
\end{equation}
then Theorem~\ref{thm:hard} applies and
\[
\bar\alpha(D)=\frac{1+\ind_{\{V(D)>0\}}}{2},
\qquad
\bRF^n(D)=nR+\bar\alpha(D)\log n+O(1).
\]
\end{corollary}

For the almost-sure result, for each $x^n\in\mathcal X^n$, let
\[
T_{x^n}(x)
=\frac1n\sum_{i=1}^n\ind_{\{x_i=x\}},
\qquad x\in\mathcal X,
\]
denote its empirical distribution.
For a sufficiently small fixed $\varepsilon>0$, define
\begin{equation}
G_n=\left\{x^n:\|T_{x^n}-P_X\|_\infty\le\varepsilon\right\},
\label{eq:finite-good-set}
\end{equation}
where $\varepsilon>0$ is chosen sufficiently small for the covering and concentration arguments below. In particular, Hoeffding's inequality gives
\begin{equation}
P_X^{\otimes n}(G_n^c)
\le 2|\mathcal X|e^{-2n\varepsilon^2}.
\label{eq:finite-good-tail}
\end{equation}
The same sets $G_n$ will be used to verify all the assumptions of
Theorem~\ref{thm:hard}.

\subsection{Dual Attainment and Average-Distortion Stability}
We first verify \eqref{eq:tilted}--\eqref{eq:dual-bound},
\eqref{eq:stability}, and \eqref{eq:average-moment}, as required
by Theorem~\ref{thm:average}. These assertions do not require
\eqref{eq:hard-feasible}.

The finite channel simplex is compact, so a rate-distortion optimizer exists. 
Since $D>D_{\min}$, strict feasibility supplies a finite Lagrange multiplier $s\ge0$. 
Since $D<D_0$ and $R>0$, this multiplier cannot be zero. Thus $s>0$, and complementary slackness shows that every optimizer has expected distortion $D$.
For a channel $W$ and $(X,Y)\sim P_XW$, recall that
\[
I(X;Y)=\min_q\sum_xP_X(x)D(W(\cdot|x)\|q).
\]
Minimizing the Lagrangian with $q^\star$ the output
marginal of an optimizer gives
\begin{align*}
W^\star(y|x)=\frac{q^\star(y)2^{-sd(x,y)}}
{\sum_zq^\star(z)2^{-sd(x,z)}}=q^\star(y)2^{j(x)+s(D-d(x,y))}.
\end{align*}
It follows that
\begin{align*}
R=\sum_{x,y}P_X(x)W^\star(y|x)
\log\frac{W^\star(y|x)}{q^\star(y)}=\E[j(X)+s(D-d(X,Y))]=\E j(X),
\end{align*}
where the sums are restricted to $q^\star(y)>0$ when necessary.

To verify the dual normalization for all $y\in\mathcal Y$, minimize the function $-\sum_xP_X(x)\log\sum_yq(y)2^{-sd(x,y)}$ over $q$.
Its partial derivative at $q^\star$ is
\begin{align*}
-\frac1{\ln2}\sum_xP_X(x)
\frac{2^{-sd(x,y)}}{\sum_zq^\star(z)2^{-sd(x,z)}} =-\frac1{\ln2}
\E_{P_X}2^{j(X)+s(D-d(X,y))}.
\end{align*}
Moreover,
\begin{align*}
\sum_yq^\star(y)
\E_{P_X}2^{j(X)+s(D-d(X,y))}=\E_{P_X}\left[
2^{j(X)}\sum_yq^\star(y)2^{s(D-d(X,y))}\right]=1.
\end{align*}
Hence this expectation equals one on $\supp q^\star$ and is at
most one for every other $y$. Since $P_X$ has full support and
the alphabets are finite, the expectation is also strictly
positive. This proves \eqref{eq:tilted}--\eqref{eq:dual-bound}.

Next, because $\mathcal Y$ is finite and
\eqref{eq:exhaustion} holds, there is an integer $m_0$ such that
\[
\mathcal Y_m=\{y:Q_y\ne P_X\},\qquad m\ge m_0.
\]
Let $W_k$ be any sequence of distortion-feasible
channels, and let $(X,Y_k)\sim P_XW_k$.
Suppose that $I(X;Y_k)\to R$. Choose a subsequence along which
$P_Y^{W_k}(\mathcal Y_{m_0})$ approaches its lower limit.
By compactness, there exists a further subsequence
converging to a channel $W$.
Let $(X,Y)\sim P_XW$.
By continuity of distortion and mutual information,
$W$ is distortion-feasible and
\[
I(X;Y)=\lim_{k\to\infty}I(X;Y_k)=R.
\]
Thus $W\in\mathcal W_D$, and
\begin{align*}
\liminf_{k\to\infty}P_Y^{W_k}(\mathcal Y_{m_0})
=P_Y^W(\mathcal Y_{m_0})\ge\min_{W'\in\mathcal W_D}
P_Y^{W'}\{y:Q_y\ne P_X\}.
\end{align*}
Since $\mathcal Y_m=\mathcal Y_{m_0}$ for $m\ge m_0$,
this proves \eqref{eq:stability}. Finally, the optimizer set is compact and the informative-output probability is continuous on it, so the infimum in \eqref{eq:alpha-characterization} is attained. For any optimizer, Lemma~\ref{lem:dual} gives $P_{X|Y=y}=Q_y$ for every output of positive probability, and each $Q_y$ has full support on the finite source alphabet. 
The conditional log-likelihood ratios are therefore bounded, which verifies \eqref{eq:average-moment}.

\subsection{Distortion-Ball Moments and Exceptional Blocks}
We now verify \eqref{eq:covering-mean},
\eqref{eq:covering-variance}, and \eqref{eq:exceptional-cost} for the sets $G_n$ in \eqref{eq:finite-good-set}. The first two bounds concern the distortion-ball probability $p_n$ in \eqref{eq:ball-prob}, and the last controls the source blocks outside $G_n$.

For a probability vector $t$ on $\mathcal X$, define
\[
F(t)=\sup_{u\ge0}\left\{
-uD-\sum_xt(x)\log\sum_yq^\star(y)2^{-ud(x,y)}
\right\}.
\]
At $t=P_X$, the maximizing multiplier is $s$ and $F(P_X)=R$.
For $u\ge0$, let
\[
W_u(y|x)=
\frac{q^\star(y)2^{-ud(x,y)}}
{\sum_zq^\star(z)2^{-ud(x,z)}}.
\]
The conditional distortion variance at $(P_X,s)$ is $\sum_xP_X(x)\Var_{W_s(\cdot|x)}d(x,Y)>0$.
If this sum were zero, then, because $P_X$ has full
support and $W_s(\cdot|x)$ is positive on $\supp q^\star$, $d(x,y)$ would be constant in $y\in\supp q^\star$ for each $x$. The formula for $W_s$ would then imply $W_s(\cdot|x)=q^\star$ for every $x$, contradicting $R>0$.

The tilted mean distortion satisfies
\[
\frac{\partial}{\partial u}
\sum_xt(x)\E_{W_u(\cdot|x)}d(x,Y)
=-(\ln2)\sum_xt(x)\Var_{W_u(\cdot|x)}d(x,Y).
\]
This derivative is strictly negative at $(P_X,s)$.
By the implicit function theorem, for $t$
sufficiently close to $P_X$, there exists a
continuously differentiable maximizing multiplier
$u(t)>0$, with $u(P_X)=s$, such that
\[
\sum_x t(x)
\E_{W_{u(t)}(\cdot|x)}d(x,Y)=D.
\]
Since $u(t)$ is continuous and $\sum_xP_X(x)\Var_{W_s(\cdot|x)}d(x,Y)>0$,
we may shrink $\varepsilon$ in
\eqref{eq:finite-good-set} so that, for every
$t$ satisfying
$\|t-P_X\|_\infty\le\varepsilon$,
\[
c_1\le
\sum_xt(x)\Var_{W_{u(t)}(\cdot|x)}d(x,Y)
\le c_2
\]
for some constants $0<c_1<c_2<\infty$.

Now fix $x^n\in G_n$ and let $t=T_{x^n}$.
For independent random variables
$Y_i\sim W_{u(t)}(\cdot|x_i)$, the definition
of the empirical distribution gives
\begin{align*}
\sum_{i=1}^n\E d(x_i,Y_i)=n\sum_xt(x)\E_{W_{u(t)}(\cdot|x)}d(x,Y)=nD,
\end{align*}
and
\begin{align*}
\sum_{i=1}^n\Var(d(x_i,Y_i))=n\sum_xt(x)
\Var_{W_{u(t)}(\cdot|x)}d(x,Y)\in[c_1n,c_2n].
\end{align*}
Furthermore, since the alphabets are finite,
the distortion function is bounded, and hence
\begin{align*}
\sum_{i=1}^n
\E\left|
d(x_i,Y_i)-\E d(x_i,Y_i)
\right|^3\le
n\left(
\max_{x,y}d(x,y)-\min_{x,y}d(x,y)
\right)^3\le c_3n
\end{align*}
for some constant $c_3<\infty$.
All three bounds hold uniformly over
$x^n\in G_n$.
By the Berry--Esseen bound~\cite{petrov1995limit}, these
estimates imply that
\[
\Pr\left\{\sum_i d(x_i,Y_i)\in[a,a+1)\right\}
\le\frac C{\sqrt n}.
\]
They also imply that, for a sufficiently large fixed $M>0$,
\[
\Pr\left\{nD-M\le\sum_i d(x_i,Y_i)\le nD\right\}
\ge\frac c{\sqrt n}.
\]

For fixed $x^n\in G_n$ of type $t$, the change of measure
from $W_{u(t)}^{\otimes n}(\cdot|x^n)$ to
$(q^\star)^{\otimes n}$ gives
\begin{align*}
\frac{(q^\star)^{\otimes n}(y^n)}
{\prod_iW_{u(t)}(y_i|x_i)}
&=2^{-nF(t)}
2^{u(t)(\sum_i d(x_i,y_i)-nD)},\\
p_n(x^n)
&=2^{-nF(t)}\E\left[
2^{-u(t)(nD-\sum_i d(x_i,Y_i))}
\ind_{\{\sum_i d(x_i,Y_i)\le nD\}}
\right],
\end{align*}
where the expectation is under the independent tilted
variables $Y_i\sim W_{u(t)}(\cdot|x_i)$.
Restricting the expectation to the event $nD-M\le\sum_i d(x_i,Y_i)\le nD$,
on which $2^{-u(t)(nD-\sum_i d(x_i,Y_i))} \ge 2^{-u(t)M}$, we obtain
\begin{align*}
p_n(x^n)
&\ge 2^{-nF(t)}2^{-u(t)M}
\Pr\left\{
nD-M\le\sum_i d(x_i,Y_i)\le nD
\right\}\\
&\ge \frac{c_0}{\sqrt n}
2^{-nF(t)}2^{-u(t)M}\\
&\ge \frac{c}{\sqrt n}2^{-nF(t)},
\end{align*}
where the second inequality follows from the preceding Berry--Esseen lower bound, and the last inequality uses the uniform boundedness of $u(t)$ on $G_n$.

For the upper bound, partition the nonnegative distortion into unit intervals. Using the uniform unit interval probability bound and $\inf_{t\in G_n}u(t)>0$, we obtain
\begin{align*}
p_n(x^n)
&\le2^{-nF(t)}\sum_{k\ge0}
2^{-ku(t)}
\Pr\left\{nD-\sum_i d(x_i,Y_i)\in[k,k+1)\right\}\\
&\le\frac C{\sqrt n}2^{-nF(t)}
\sum_{k\ge0}2^{-ku(t)}\\
&\le\frac{C'}{\sqrt n}2^{-nF(t)}.
\end{align*}
Thus, for all $x^n\in G_n$,
\begin{equation}
\frac c{\sqrt n}2^{-nF(T_{x^n})}
\le p_n(x^n)\le
\frac C{\sqrt n}2^{-nF(T_{x^n})}.
\label{eq:finite-covering}
\end{equation}

The derivative of $F$ at $P_X$ is $j(x)+sD$. 
Since $u(t)$ is continuously differentiable near $P_X$ and $u(P_X)=s$, there exists a constant $C<\infty$ such that
\[
|u(t)-s|\le C\|t-P_X\|_\infty.
\]
Moreover, the objective defining $F(t)$ has zero derivative at its maximizer $u(t)$ and a uniformly bounded second derivative in a neighborhood of $s$.
Taylor's theorem therefore yields
\begin{align*}
0 \le F(t)-\sum_xt(x)j(x) \le C|u(t)-s|^2 \le C\|t-P_X\|_\infty^2.
\end{align*}
Taking $t=T_{x^n}$ and using
$n\sum_xt(x)j(x)=\sum_{i=1}^n j(x_i)$,
we obtain
\begin{equation}
0\le nF(T_{x^n})-\sum_{i=1}^n j(x_i)
\le Cn\|T_{x^n}-P_X\|_\infty^2,
\qquad x^n\in G_n.
\label{eq:finite-taylor}
\end{equation}
It follows from \eqref{eq:finite-covering} that
\[
\left|-
\log p_n(x^n)-\sum_i j(x_i)-\frac12\log n\right|
\le C\left(1+n\|T_{x^n}-P_X\|_\infty^2\right),
\qquad x^n\in G_n.
\]
Since $X_1,\ldots,X_n$ are i.i.d., for every
$x\in\mathcal X$,
\[
\begin{aligned}
\E|T_{X^n}(x)-P_X(x)|^2
&=\frac{P_X(x)(1-P_X(x))}{n}
\le\frac1{4n},\\
\E|T_{X^n}(x)-P_X(x)|^4
&=\frac1{n^4}\Big[
n\E|\ind_{\{X_1=x\}}-P_X(x)|^4+
3n(n-1)P_X(x)^2(1-P_X(x))^2
\Big]\\
&\le\frac1{n^3}+\frac3{16n^2}.
\end{aligned}
\]
Consequently, since $\mathcal X$ is finite,
\begin{align*}
\E\|T_{X^n}-P_X\|_\infty^2
&\le\sum_{x\in\mathcal X}
\E|T_{X^n}(x)-P_X(x)|^2=O(n^{-1}),\\
\E\|T_{X^n}-P_X\|_\infty^4
&\le\sum_{x\in\mathcal X}
\E|T_{X^n}(x)-P_X(x)|^4=O(n^{-2}).
\end{align*}
Therefore,
\begin{align*}
\E\left[
\left(-\log p_n(X^n)-\sum_i j(X_i)
-\tfrac12\log n\right)^2
\ind_{\{X^n\in G_n\}}\right]\le C\left(1+n^2
\E\|T_{X^n}-P_X\|_\infty^4\right)
=O(1).
\end{align*}

Taking expectations and using
$\E\|T_{X^n}-P_X\|_\infty^2=O(n^{-1})$,
we obtain
\begin{align*}
&\E[-\log p_n(X^n)\ind_{\{X^n\in G_n\}}]\\
&\quad\le
\E\left[\sum_{i=1}^n j(X_i)
\ind_{\{X^n\in G_n\}}\right]
+\frac12\log n+Cn\E\left[
\|T_{X^n}-P_X\|_\infty^2
\ind_{\{X^n\in G_n\}}\right]+O(1)\\
&\quad\le
\E\left[\sum_{i=1}^n j(X_i)
\ind_{\{X^n\in G_n\}}\right]
+\frac12\log n
+Cn\E\|T_{X^n}-P_X\|_\infty^2+O(1)\\
&\quad=
\E\left[\sum_{i=1}^n j(X_i)
\ind_{\{X^n\in G_n\}}\right]
+\frac12\log n+O(1)\\
&\quad=
nR-\E\left[\sum_{i=1}^n j(X_i)
\ind_{\{X^n\notin G_n\}}\right]
+\frac12\log n+O(1)\\
&\quad\le
nR+n\|j\|_\infty
P_X^{\otimes n}(G_n^c)
+\frac12\log n+O(1)\\
&\quad=
nR+\frac12\log n+O(1),
\end{align*}
where the last equality follows from
\eqref{eq:finite-good-tail}.
This verifies \eqref{eq:covering-mean}.

Furthermore,
\begin{align*}
\Var(\log p_n(X^n)\mid X^n\in G_n)&\le
2\Var\left(\sum_i j(X_i)\mid X^n\in G_n\right)\\
&\quad +2\E\left[\left(
\log p_n(X^n)+\sum_i j(X_i)+\tfrac12\log n
\right)^2\middle|X^n\in G_n\right]\\
&\le
\frac{2nV(D)}{P_X^{\otimes n}(G_n)}+O(1)=
\begin{cases}
O(1),&V(D)=0,\\
O(n),&V(D)>0.
\end{cases}
\end{align*}
Here the conditional variance of $\sum_i j(X_i)$ was
bounded by its conditional squared deviation from $nR$.
This proves \eqref{eq:covering-variance}.

Finally, condition \eqref{eq:hard-feasible} allows us to choose
for every $x$ a reproduction letter $y(x)$ satisfying
$d(x,y(x))\le D$. For $x^n\notin G_n$, reproduce
$(y(x_1),\ldots,y(x_n))$ and encode this block by a fixed-length
prefix description with
$\ell_n(x^n)\le n\lceil\log|\mathcal Y|\rceil$.
The distortion constraint holds coordinatewise, and
\begin{align*}
&\E\left[\left(
\left|\sum_i j(X_i)\right|+\log n+\ell_n(X^n)
\right)\ind_{\{X^n\notin G_n\}}\right]\le\left(n\|j\|_\infty+\log n
+n\lceil\log|\mathcal Y|\rceil\right)
P_X^{\otimes n}(G_n^c)=O(1).
\end{align*}
This proves \eqref{eq:exceptional-cost}.

\subsection{Local Concentration}

We verify \eqref{eq:hard-local} for the sets $G_n$ in
\eqref{eq:finite-good-set}. When $V(D)=0$, a scalar
Kolmogorov--Rogozin bound suffices
\cite{rogozin1961increase}; when $V(D)>0$, we need
a two-dimensional concentration bound. We first prove
the following finite-alphabet version of a standard
type of multidimensional anti-concentration estimate
\cite{larin1994concentration}.

\begin{lemma}\label{lem:two-dimensional}
Let $\{\mu_y:y\in\mathcal Y\}$ be a fixed finite family
of finitely supported distributions on $\R^2$, and let
$A_i\sim\mu_{y_i}$ be independent. If
\[
\frac1n\sum_{i=1}^n\Cov(A_i)\succeq\kappa I_2
\qquad\text{for some }\kappa>0,
\]
then
\[
\sup_{z\in\R^2}\Pr\left\{
\sum_{i=1}^n A_i\in z+[0,1)^2\right\}
\le\frac{C_\kappa}{n},
\]
where $C_\kappa$ may also depend on the fixed family.
\end{lemma}

\begin{proof}
Choose $\delta>0$ sufficiently small that
$\delta\sum_y\Cov_{\mu_y}(A)\preceq
(\kappa/2)I_2$. Discard the symbols occurring
fewer than $\delta n$ times in $y^n$.
Because covariance matrices are positive semidefinite,
the remaining symbols satisfy
\begin{align*}
&\sum_{\substack{y\in\mathcal Y:\\
\sum_i\ind_{\{y_i=y\}}\ge\delta n}}
\frac{\sum_i\ind_{\{y_i=y\}}}{n}\Cov_{\mu_y}(A)\succeq
\frac1n\sum_{i=1}^n\Cov(A_i)
-\delta\sum_{y\in\mathcal Y}\Cov_{\mu_y}(A)\succeq\frac\kappa2 I_2.
\end{align*}
The identity
\[
\Cov_{\mu_y}(A)
=\frac12\sum_{a,b\in\supp\mu_y}
\mu_y(a)\mu_y(b)(a-b)(a-b)^{\mathsf T}
\]
then shows that the differences between support points
of the remaining distributions span $\R^2$.
Hence two such differences, $a_1-b_1$ and $a_2-b_2$,
are linearly independent, and their distributions
each occur in at least $\delta n$ coordinates.
If both differences come from the same distribution,
we divide its coordinates into two disjoint groups;
otherwise, use the two groups of corresponding
coordinates. In either case, each group has at least
$cn$ coordinates for some $c>0$ independent of $n$.

Each selected distribution assigns positive mass to
both support points $a_j$ and $b_j$. It can therefore
be sampled by first choosing, with probability
\[
2\min\{\mu_{y^{(j)}}(a_j),\mu_{y^{(j)}}(b_j)\}>0,
\]
a uniform draw from $\{a_j,b_j\}$, and otherwise
drawing from the remaining probability distribution.
Here $y^{(j)}$ is the symbol associated with the
$j$th group. Because the family is finite, the
probabilities of selecting these uniform draws
are bounded away from zero. Chernoff's inequality
shows that, except with probability $Ce^{-cn}$,
each group contains at least $cn$ such draws.

Condition on which coordinates use uniform draws,
on all remaining outcomes, and on the coordinates
outside the two groups. The sum of the $A_i$ is
then a fixed translation of
\[
(a_1-b_1)B_1+(a_2-b_2)B_2,
\]
where $B_1,B_2$ are independent binomial variables
with parameters $(m_1,1/2)$ and $(m_2,1/2)$,
respectively, and $m_1,m_2\ge cn$.
The standard binomial bound gives
\[
\sup_{k,l\in\mathbb Z}\Pr\{B_1=k,B_2=l\}
\le\frac{C}{\sqrt{m_1m_2}}\le\frac{C'}n.
\]
Since $a_1-b_1$ and $a_2-b_2$ are linearly
independent, a translated unit square contains
at most a fixed number of points of the lattice
they generate. The number can be bounded uniformly
because there are only finitely many possible
differences between support points. Thus,
conditionally on the event above, the probability
of any translated unit square is at most $C/n$.
Removing the conditioning and using the Chernoff
bound yields
\[
\sup_{z\in\R^2}\Pr\left\{
\sum_{i=1}^n A_i\in z+[0,1)^2\right\}
\le\frac Cn+Ce^{-cn}\le\frac{C_\kappa}{n}.
\]
\end{proof}

Suppose first that $V(D)=0$. Since $P_X$ has
full support, $j(x)=R$ for every $x$. If a distortion
column $d(x,y)$ is constant in $x$, then
\eqref{eq:dual-bound} gives
\[
1\ge\E_{P_X}2^{j(X)+s(D-d(X,y))}
=2^{R+s(D-d(x,y))},
\]
so $d(x,y)\ge D+R/s$ for every $x$.
For any $y^n$ with a nonempty distortion ball,
let $k$ be the number of its nonconstant distortion
columns. For a source block in that ball,
\begin{align*}
nD\ge\sum_{i=1}^n d(x_i,y_i)\ge(n-k)(D+R/s),
\end{align*}
which implies $k\ge nR/(sD+R)$. Since $\mathcal Y$
is finite, one nonconstant column occurs at least
$cn$ times. Under $\nu_{y^n}$, the corresponding
distortions are independent copies of a nonconstant
random variable, because $Q_y$ has full support.
The Kolmogorov--Rogozin inequality therefore gives
\[
\sup_{v\in\R}\nu_{y^n}\left\{
nD-\sum_{i=1}^n d(X_i,y_i)\in[v,v+1)
\right\}\le\frac C{\sqrt n}.
\]
Since $\sum_i j(X_i)=nR$, this also bounds the
event in \eqref{eq:hard-local}. If the distortion
ball is empty, that event has probability zero
for $v\ge0$.

Now suppose $V(D)>0$, so $j$ is nonconstant.
We first show that for every probability vector
$q$ on $\mathcal Y$ satisfying
\begin{equation}
\sum_yq(y)Q_y=P_X,\qquad
\sum_yq(y)\E_{Q_y}d(X,y)\le D,
\label{eq:finite-marginal}
\end{equation}
the matrix
\[
\sum_yq(y)\Cov_{Q_y}
\begin{pmatrix}j(X)\\d(X,y)\end{pmatrix}
\]
is positive definite. 
We prove the positive definiteness by contradiction.
Suppose that the covariance matrix is singular.
Then there exists $(u,v)\ne(0,0)$ such that
\[
\sum_y q(y)\Var_{Q_y}
\bigl(uj(X)+vd(X,y)\bigr)=0.
\]
Since $j$ is nonconstant and every $Q_y$ has
full support, we must have $v\ne0$.
Moreover, since each variance is nonnegative,
\[
u(j(x)-j(x'))+
v(d(x,y)-d(x',y))=0
\]
for all $x,x'\in\mathcal X$ and $y$ with $q(y)>0$.

By \eqref{eq:dual-q}, we obtain
\begin{align*}
\log\frac{Q_y(x)P_X(x')}{Q_y(x')P_X(x)}=j(x)-j(x')
-s(d(x,y)-d(x',y))=\left(1+\frac{su}{v}\right)
(j(x)-j(x')).
\end{align*}
Since the right-hand side is independent of $y$,
all $Q_y$ with $q(y)>0$ are identical.
The first condition in \eqref{eq:finite-marginal}
therefore implies $Q_y=P_X$ whenever $q(y)>0$.
Consequently, for every such $y$,
\begin{align*}
0=\E_{P_X}\log\frac{Q_y(X)}{P_X(X)}=R+s(D-\E_{P_X}d(X,y))-\log\E_{P_X}
2^{j(X)+s(D-d(X,y))}\ge R+s(D-\E_{P_X}d(X,y)),
\end{align*}
where the last inequality follows from
\eqref{eq:dual-bound}. Hence, $\E_{Q_y}d(X,y) =\E_{P_X}d(X,y) \ge D+\frac Rs$.
Averaging over $q$ gives
\begin{align*}
\sum_yq(y)\E_{Q_y}d(X,y)\ge\sum_yq(y)\left(D+\frac Rs\right)=D+\frac Rs>D,
\end{align*}
contradicting the second condition in
\eqref{eq:finite-marginal}.
Therefore, the covariance matrix is
positive definite.

We next prove a lower bound
on the smallest eigenvalue by contradiction. Suppose that no such bound exists.
Then, for every positive integer $k$, there is
a probability vector $q_k$ satisfying
\[
\left\|\sum_yq_k(y)Q_y-P_X\right\|_\infty
\le\frac2k,\qquad
\sum_yq_k(y)\E_{Q_y}d(X,y)
\le D+\frac2k,
\]
but
\[
\lambda_{\min}\left(
\sum_yq_k(y)\Cov_{Q_y}
\begin{pmatrix}
j(X)\\d(X,y)
\end{pmatrix}
\right)<\frac1k.
\]
Since the probability simplex is compact,
a subsequence of $q_k$ converges to some
probability vector $q$. By continuity,
\[
\sum_yq(y)Q_y=P_X,\qquad
\sum_yq(y)\E_{Q_y}d(X,y)\le D,
\]
and
\[
\lambda_{\min}\left(
\sum_yq(y)\Cov_{Q_y}
\begin{pmatrix}
j(X)\\d(X,y)
\end{pmatrix}
\right)=0.
\]
This contradicts the positive definiteness
proved above. 

First, suppose we can choose $\varepsilon>0$ such that \begin{equation}
\left\|\sum_y q(y)Q_y-P_X\right\|_\infty
\le 2\varepsilon,
\qquad
\sum_y q(y)\E_{Q_y}d(X,y)
\le D+2\varepsilon,
\label{eq:finite-covariance}
\end{equation}
implying \[
\lambda_{\min}\left(
\frac1n\sum_i\operatorname{Cov}_{Q_{y_i}}
\begin{pmatrix}j(X)\\d(X,y_i)\end{pmatrix}
\right)\ge\kappa.
\]
Then fix $y^n$ and apply Lemma~\ref{lem:two-dimensional}
with
\[
A_i=
\begin{pmatrix}
j(X_i)\\
-d(X_i,y_i)
\end{pmatrix},
\qquad X_i\sim Q_{y_i},
\]
and
\[
\mu_y=
\sum_{x\in\mathcal X}Q_y(x)
\delta_{(j(x),-d(x,y))},
\qquad y\in\mathcal Y.
\]
Under $\nu_{y^n}=\bigotimes_{i=1}^n Q_{y_i}$,
the vectors $A_i$ are independent and
satisfy $A_i\sim\mu_{y_i}$. Moreover,
$\{\mu_y:y\in\mathcal Y\}$ is a fixed finite
family of finitely supported distributions
on $\R^2$.

Applying \eqref{eq:finite-covariance} to
the empirical frequencies of $y^n$ gives
\[
\frac1n\sum_{i=1}^n
\Cov_{Q_{y_i}}
\begin{pmatrix}
j(X)\\d(X,y_i)
\end{pmatrix}
\succeq\kappa I_2.
\]
Consequently,
\begin{align*}
\frac1n\sum_{i=1}^n\Cov(A_i)=
\begin{pmatrix}1&0\\0&-1\end{pmatrix}
\left(
\frac1n\sum_{i=1}^n
\Cov_{Q_{y_i}}
\begin{pmatrix}
j(X)\\d(X,y_i)
\end{pmatrix}
\right)
\begin{pmatrix}1&0\\0&-1\end{pmatrix}\succeq\kappa I_2.
\end{align*}
Thus, all the conditions of
Lemma~\ref{lem:two-dimensional} are satisfied.
Since
\[
\sum_{i=1}^n A_i
=
\begin{pmatrix}
\sum_i j(X_i)\\
-\sum_i d(X_i,y_i)
\end{pmatrix},
\]
the lemma gives
\begin{align*}
\sup_{u\in\R,\,v\ge0}
\nu_{y^n}\left\{
\begin{array}{l}
\sum_i j(X_i)\in[u,u+1),\\
nD-\sum_i d(X_i,y_i)\in[v,v+1)
\end{array}
\right\}&=
\sup_{u\in\R,\,v\ge0}
\nu_{y^n}\left\{
\sum_{i=1}^n A_i
\in (u,v-nD)+[0,1)^2
\right\}\\
&\le
\sup_{z\in\R^2}
\nu_{y^n}\left\{
\sum_{i=1}^n A_i\in z+[0,1)^2
\right\}\le\frac{C_\kappa}{n}.
\end{align*}
Restricting the event to $X^n\in G_n$
preserves this upper bound.

Suppose that at least one of the two conditions in \eqref{eq:finite-covariance} fails.
If
\[
\left\|\frac1n\sum_{i=1}^n Q_{y_i}
-P_X\right\|_\infty>2\varepsilon,
\]
then, for $X^n\in G_n$,
\begin{align*}
\left\|T_{X^n}
-\frac1n\sum_{i=1}^n Q_{y_i}\right\|_\infty\ge
\left\|\frac1n\sum_{i=1}^nQ_{y_i}
-P_X\right\|_\infty-\|T_{X^n}-P_X\|_\infty>\varepsilon.
\end{align*}
Since $X_i\sim Q_{y_i}$ independently under
$\nu_{y^n}$, Hoeffding's inequality gives
\begin{align*}
\nu_{y^n}\{X^n\in G_n\}\le
\sum_{x\in\mathcal X}
\nu_{y^n}\left\{
\left|
\frac1n\sum_{i=1}^n
\bigl(\ind_{\{X_i=x\}}-Q_{y_i}(x)\bigr)
\right|>\varepsilon
\right\}\le 2|\mathcal X|e^{-2n\varepsilon^2}.
\end{align*}

Otherwise, the second condition must fail, namely,
\[
\frac1n\sum_{i=1}^n
\E_{Q_{y_i}}d(X,y_i)>D+2\varepsilon.
\]
In this case, the distortion constraint
$d_n(X^n,y^n)\le D$ implies
\[
\frac1n\sum_{i=1}^n
\left(\E_{Q_{y_i}}d(X,y_i)-d(X_i,y_i)\right)
>2\varepsilon.
\]
Since the distortion function is bounded
and the variables are independent,
Hoeffding's inequality yields
\begin{align*}
\nu_{y^n}\{d_n(X^n,y^n)\le D\}\le
\nu_{y^n}\left\{
\frac1n\sum_{i=1}^n
\left(\E_{Q_{y_i}}d(X,y_i)-d(X_i,y_i)\right)
>2\varepsilon
\right\}\le Ce^{-cn}.
\end{align*}

Thus, in either case,
\begin{align*}
\nu_{y^n}\left\{
X^n\in G_n,\,
d_n(X^n,y^n)\le D
\right\}\le Ce^{-cn}\le\frac{C'}n.
\end{align*}
Consequently, for every $u\in\R$ and $v\ge0$,
\begin{align*}
\nu_{y^n}\left\{
\begin{array}{l}
X^n\in G_n,\\
\sum_i j(X_i)\in[u,u+1),\\
nD-\sum_i d(X_i,y_i)\in[v,v+1)
\end{array}
\right\}\le
\nu_{y^n}\left\{
X^n\in G_n,\,
d_n(X^n,y^n)\le D
\right\}\le\frac{C'}n.
\end{align*}
Combining this with the bound obtained from
Lemma~\ref{lem:two-dimensional} when both
conditions in \eqref{eq:finite-covariance}
hold, we verify \eqref{eq:hard-local} uniformly
over $y^n$, $u$, and $v$ when $V(D)>0$.
This completes the finite-alphabet verification.

\section*{Acknowledgment}
The author sincerely thanks Professor Aaron B.~Wagner
for bringing this research direction to the author's attention and for his extensive participation in technical discussions throughout the development
of this work. 
His insights and suggestions
have contributed substantially to the results presented in this paper.

\newpage
\bibliographystyle{IEEEtran}
\bibliography{ref}

\end{document}